# Modified silicone rubbers for fabrication and contacting of flexible suspended membranes of n-/p-GaP nanowires with single-walled carbon nanotube transparent contact

Vladimir Neplokh [1*], Fedor M. Kochetkov [1], Konstantin V. Deriabin [2], Vladimir V. Fedorov [1], Alexey D. Bolshakov [1], Igor E. Eliseev [1], Vladimir Yu. Mikhailovskii [3], Daniil A. Ilatovskii [4], Dmitry V. Krasnikov [4], Maria Tchernycheva [6], George E. Cirlin [1], Albert G. Nasibulin [4, 5], Ivan S. Mukhin [1, 7], Regina M. Islamova [2**]

[1] St. Petersburg Academic University, Khlopina 8/3, 194021, Saint Petersburg, Russia
[2] Institute of Chemistry, Saint Petersburg State University, 7/9 Universitetskaya Nab., 199034 Saint Petersburg, Russia
[3] Saint Petersburg State University, 13B Universitetskaya Emb., St Petersburg 199034, Russia
[4] Skolkovo Institute of Science and Technology, Nobel str. 3, Moscow, 121205, Russia
[5] Aalto University School of Chemical Engineering, Kemistintie 1, 02015, Espoo, Finland
[6] Centre of Nanosciences and Nanotechnologies, UMR 9001 CNRS, Univ. Paris Sud, Univ. Paris-Saclay, 10 Boulevard Thomas Gobert, 91120 Palaiseau Cedex, France
[7] ITMO University, Kronverksky 49, 197101, Saint Petersburg, Russia

*vneplox@gmail.com +7 9052553999, **r.islamova@spbu.ru +7 9117294724

1. Abstract

This work proposes new chemical and mechanical materials and techniques for III-V semiconductor NW/silicone membrane formation and optoelectronic device fabrication. Molecular beam epitaxy (MBE)-synthesized n-, p- and i-GaP NWs were encapsulated by introduced G-coating method into synthesized polydimethylsiloxane-graft-polystyrene and released from the Si growth substrate. The fabricated membranes were contacted with different materials including single-walled carbon nanotubes or ferrocenyl-containing polymethylhydrosiloxane with and without multi-walled carbon nanotubes doping. The electrical connection of the fabricated membranes was verified by electron beam induced current (EBIC) spectroscopy. The developed methods and materials can be applied for fabrication of high quality flexible inorganic optoelectronic devices.

2. Introduction

The appealing properties of organic light emitting diodes (OLEDs), i. e. relatively easy and inexpensive fabrication, and efficient electroluminescence (EL) allowed the OLED-based industry to conquer a significant market share. For instance, modern smartphones are mostly produced with the OLED displays [weblink1, weblink2]. However, organic materials are far behind the inorganic materials in terms of stability and external quantum efficiency (EQE) of EL in optical range, especially in blue and red region [Sal2019, Mon2019], which for inorganic devices becomes close to 100%. Inorganic LEDs based on compounds of arsenides, nitrides, phosphides etc. are envisioned to be the materials for the LEDs with the efficiency close to the theoretical limit. The recent commercial application of OLEDs instead of inorganic materials is explained mainly by difficulties of combination of different radiative materials necessary for an RGB full color screen. Indeed, the mainstream thin film technology is hard to adapt for small high resolution screen, because it requires either advanced post-growth processing, or combination of very different crystalline materials [Kle2013]. The flexible devices fabrication based on thin films imposes even greater complications, i. e. ultra thin wafer epitaxy or release of the synthesized material from the wafer [Park2009].

Nanowire (NW) or microwire (MW) design of inorganic devices has several significant advantages, especially for substrate-free device fabrication. Wires have a small footprint, therefore they can be mechanically removed from the initial growth substrate [Nad2008, Dai2015]. High surface to volume ratio leads to an effective relaxation of the elastic strain due to the lattice mismatch of III-V heterostructures, therefore low structural defect concentration can be achieved even for a high lattice mismatch [Her2011]. Core-shell wire heterostructures also have effective light extraction and current injection [Tch2014], which is very important for optoelectronic applications, e. g. LEDs. One of the main attractive features of NW devices is the possibility to combine very different materials, e. g. nitrides and arsenides or phosphides. The NWs can be encapsulated into a polymer matrix (i.e. elastomer) [Lee2017, Amj2014] and then released from the wafer. The elastomer/NW membrane can then be electrically connected with conductive transparent electrodes [Dai2015], and the fabricated membranes of different materials may be stacked onto each other to form composite device with multiple line EL. The pixel contacts to membranes with different color channels could be provided independently, thus the elastomer/NW devices can be considered as inorganic analogue to OLED devices.

Despite the high attractiveness, the early demonstrations of membrane devices based on encapsulated wires did not result in effective device fabrication and usually researchers do not show significant improvement in following papers [Park2009, Nad2008]. The main reason is essentially the difficulties in fabrication and functionalization of NW devices, which are even more pronounced for membrane devices. Indeed, compared to a thin film device the elastomer/NW LED device is basically an array of billions of independently operating NWs LEDs, which requires complicated contacting to achieve a high yield, i. e. the radiating/dim NW ratio. The yield is typically low because an LED works in the steep region of I-V curve, and therefore a small variation of series resistance and material composition due to the inhomogeneous distribution of NW parameters, i. e. voltage applied to individual NWs, leads to a significant variation of current [Nep2016]. Membrane devices also have a problem of mechanical stability of contacts, and elastomer materials prevent high temperature annealing, required to achieve the ohmic resistance in many semiconductor-metal material systems [Han1993].

In order to find a solution to the problems of membrane device fabrication, we developed encapsulation and release technique suitable for few μm thick membranes, and tested different contacting strategies. The crucial point of our research is the application of advanced silicone chemistry for the specific problems of elastomer/NW membrane fabrication, because usually researchers in this field focus only on the physical side of the problems and employ very basic and widely commercially available materials, mostly Dow Corning Sylgard 182 or 184 [Li2019], which are not designed for thin membrane fabrication. Typical polydimethylsiloxane (PDMS) rubbers including Dow Corning Sylgard 182 or 184 [Li2019] are not very strong (tensile strength $\sigma$ = 0.02–0.25 MPa) [Isl2016, Isl2017, Dob2019, Der2018] and very tacky to substrates that makes them unsuitable to produce very thin and nonadhesive membranes. The use of copolymers with different vinyl monomers can rectify the problems related to silicone durability and adhesion. Thus, cross-linked graft copolymers of PDMS and polystyrene have improved tensile strength ($\sigma$ = 2.9–4.8 MPa) [Mas2017] and low adhesion that allow an easy release from the initial growth substrate.

For further development of elastomer/NW device fabrication, advanced transparent flexible conductive electrodes are required. Different strategies were introduced in literature, including thin and flexible ITO [Park2016], semitransparent metals, silver nanowires and graphene [Lee2013], carbon nanotubes (CNT) [Sun2011] etc. In this paper, we focus on single wall CNT (SWCNT) and advanced ferrocenyl-containing polymethylhydrosiloxane (FPS) as contacting layer. This contact can be further improved with diluted multi wall CNTs (MWCNTs), having high in-plane conductivity. The development of curable silicone (co)polymers with electroactive fragments in their structure is highly perspective for electrostatic discharge (ESD) protective materials for electrostatic-sensitive devices and power lines [Der2019, Wyp2016], active layers for modern sensing electronic devices [Chen2017, Wang2017, Wang2016] and biomedical implants [Kaur2015, Yang2015]. Thus, the integration of metals into a polysiloxane chain can directly influence its electronic and optical properties [Der2019, Pit2005, Abd2007]. For instance, self-curable polysiloxanes with redox-active ferrocenyl moieties were synthesized [Der2019]. They are attractive materials primarily due to the possibility of self-cross-linking, the application as a liquid contact (which is mechanically reliable and easy for use) as well as the ability to combine ferrocenyl-containing silicone rubbers with other materials, such as multi-walled nanotubes, in order to achieve high lateral conductivity.

This work is devoted to the new chemical and mechanical materials development for III-V semiconductor NW/PDMS-st membrane fabrication. The introduced G-coating method of NW encapsulation into synthesized polydimethylsiloxane-graft-polystyrene allowed an easy release of NW membrane from the Si growth substrate. Different materials including single-walled carbon nanotubes, ferrocenyl-containing polymethylhydrosiloxane with and without multi-walled carbon nanotubes doping were used to contact NWs membranes.

3. Methods

Epitaxial arrays of GaP NWs were synthesized by self-catalyzed vapor-liquid-solid (VLS) mechanism in solid source molecular beam epitaxy (MBE) process using Veeco GEN-III MBE machine. Valved phosphorus cracker was used to produce $P_2$ molecular flux at cracking temperature of 900° C. Substrate temperature was controlled with a thermocouple, calibrated using temperatures of Si(111) 7x7 ↔ 1x1 phase transitions as a references [Tel1985]. To monitor the group-III and -V element fluxes, beam equivalent pressure (BEP) was measured with the conventional Bayard–Alpert vacuum gauge. Stoichiometric P/Ga flux ratio of 6 was evaluated during the growth of planar GaP epilayers on Si(001), as the lower BEP ratios led to the accumulation of Ga droplets on sample surface [Bol2019]. Prior to the growth, the wafer was treated by the modified Shiraki cleaning procedure ended by the wet-chemical oxidation in a boiling azeotrope 68% HNO$_3$ water solution for 5 minutes at constant boiling point of 120°C [Imam2010], [Ish1986] resulting in the formation of a thin surface oxide layer acting as a growth mask for NW nucleation [Mat2015], [Bol2019b]. After thermal degassing under UHV-conditions in MBE load lock and buffer chambers, oxidized Si(111) substrates were annealed under 790°C for 30 minutes in order to create defects in the oxide layer promoting formation of catalytic Ga droplets needed for the self-catalytic VLS NW growth [Dub2012]. NW growth was started by simultaneous opening of Ga and P shutters. To obtain n- and p-type conductivity, NWs were intentionally doped by introducing the silicon or beryllium flux during the growth, respectively. It was shown, that despite the amphoteric nature of Si dopant in III-V compounds commonly n-type conductivity is observed in III-phosphide alloy NWs [Tch2012, Sak2019]. Si-doped NWs were grown at a substrate temperature ($T_{growth}$) of 640 °C and V/III ratio of 24. According to our observations introduction of Si-flux do not interrupt VLS growth only slightly affecting NW aspect ratio, while the solution of Be in Ga catalytic droplet affects its morphology and leads to the sidewalls wetting [Zhang2017]. Thus, to keep stable VLS growth of Be-doped NW, the growth temperature was increased by 10 °C, while V/III-flux ratio was decreased to 8. Growth was ended by closing Ga shutter and cooling the sample at 30°C/min under the group-V flux until 400°C [Fed2018].

Polydimethylsiloxane-graft-polystyrene (PDMS-St) was synthesized in accordance with procedure published in [Mas2017] (Figure 1). Freshly distilled styrene, $\alpha,\omega$-bis(trivinylsiloxy)polydimethyldisiloxane (published in [Mas2017]), azobisisobutyronitrile (AIBN) and ethanol were loaded into a three-necked flask equipped with a reflux condenser, stirrer, and heater. The styrene loading was 40 wt.% of $\alpha,\omega$-bis(trivinylsiloxy)polydimethyldisiloxane, and the AIBN loading was 0.8 wt.% of the total reaction mass. The resulting mixture was stirred for 4–5 h at 60–65 °C under argon atmosphere. The initiator AIBN was then decomposed at 80 °C for 2 h, and the solvent together with unreacted styrene were distilled off under reduced pressure (3–5 mm Hg) at 100–110 °C. Synthesized PDMS-St was fully characterized by NMR spectroscopy [Mas2017]. Yield of PDMS-St: 90%; white viscous liquid; viscosity 78 P.

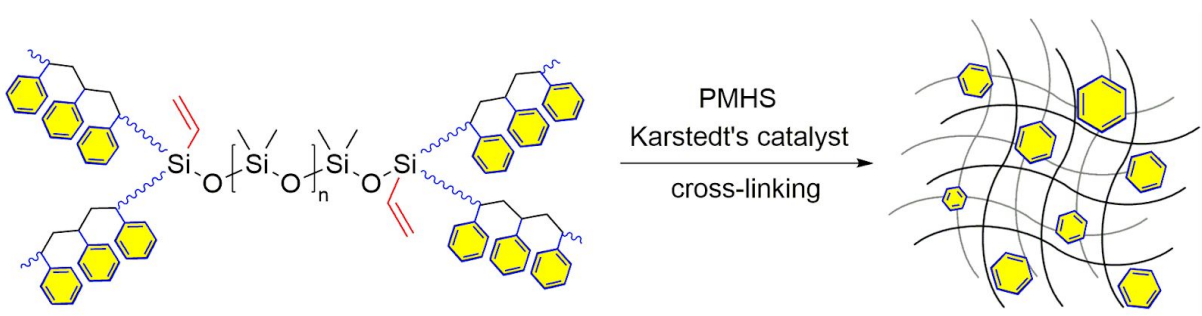

Figure 1. PDMS-St and cross-linking.

Cross-linking system of PDMS-St consists in two components. Component A includes the calculated amount of the Karstedt's catalyst solution in vinyl-terminated PDMS (0.1 M), which was mixed with PDMS-St and stirred to obtain the required concentration (2.0·10$^{-4}$ M).

Component B (cross-linker): polymethylhydrosiloxane (PMHS, viscosity 0.12–0.45 P) and PDMS-St (1:3 mass ratio) were mixed and carefully stirred. The required amounts were calculated for the specific ratio of Si–H and vinyl groups (3:1) in the reaction mixture. 0.5 mL of the component B was added to 0.5 mL of the component A and stirred for 1 min. The mixture was then placed into a desiccator at room temperature until a dry cured product was obtained. The total catalyst concentration in the final silicon rubber samples was $1.0 \cdot 10^{-4}$ M.

The common method to encapsulate NW arrays into polymer matrix is spin-coating [Pla2009, Dai2015], i. e. polymer drop-casting followed by thinning the film in vertical centrifuge similar to the photoresist application routine. The spin-coating method allows to achieve a good encapsulation of long (more than 20 µm) and low density (less than 0.1 NW per sq. µm) NW arrays, however, short and/or dense NW array embedding is challenging due to the high PDMS viscosity. This viscosity can be reduced by diluting PDMS with methylene chloride [Pla2009], toluene [Opo2015], hexane [Wei2012] or other solvents, allowing spin-casting of thin (thinner than 3 µm) PDMS films [Lot1997]. In this paper we propose a different approach for NW array encapsulation with the use of swinging bucket centrifuge, where the thinning force is perpendicular to the sample surface (Figure 2). The relative centrifugal force in bucket centrifuge can be higher compared to commonly used spin-coaters and reach 5-10 thousand G-force for standard swinging bucket rotors and up to 1 million G-force in ultracentrifugal rotors [weblink3]. For convenience and to underline similarity with gravity force we propose to call this method G-coating by analogy with spin-coating. The advantage of G-coating for the NW embedding is the high pressure applied to PDMS-St, which fills the space among the NWs.

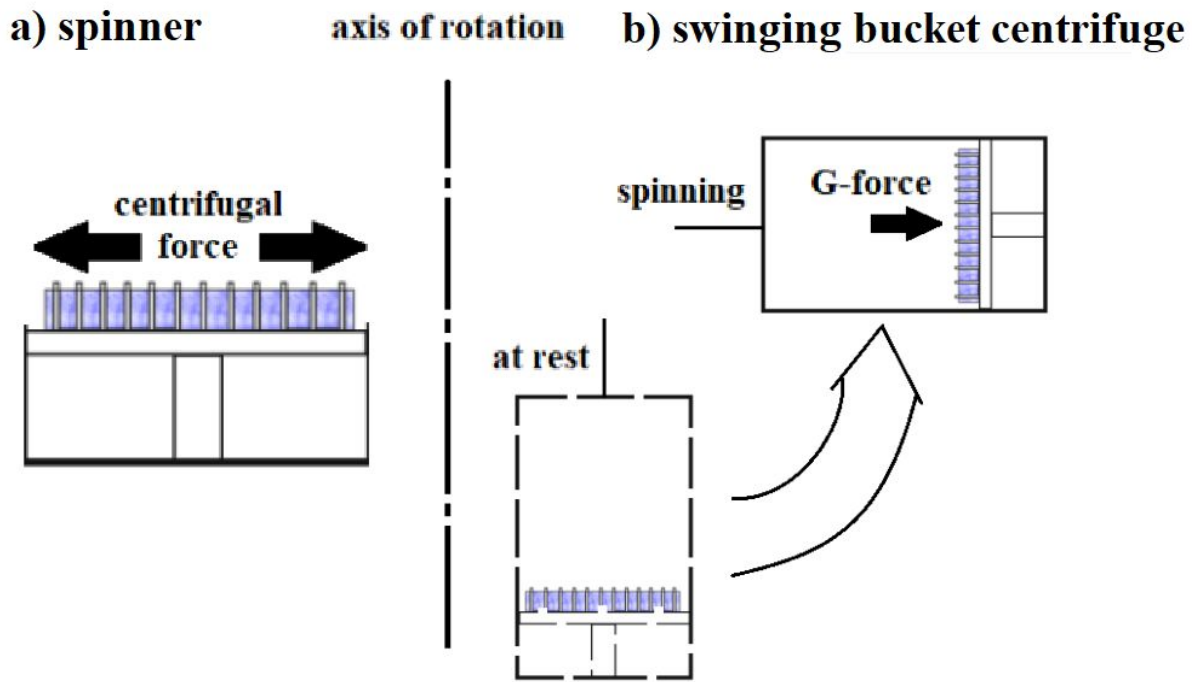

Figure 2. Spin-coating in spinner (a) and G-coating in swinging bucket centrifuge (b).

The PDMS-St/NW fabrication started with PDMS-St (component A) and cross-linker (component B) mixing 1 : 1 mass ratio, followed by debubbling in the exicator for 30-40 min. Then the prepared PDMS-St mixture was dropped onto the samples and G-coated at approx.

4500 G-force for 60 min. until the sample surface turns matt due to the light scattering by revealed NW top parts. After PDMS-St deposition, the samples were cross-linked in the oven at 80°C for 2 hours or during the night. The prepared PDMS-St/NW structures were etched in 5 cycles 40 s etching / 60 s interruption sequence for cooling, mixture of 15 and 40 ml per min. flux of $O_2$ and $CF_4$, respectively, and 150 mW RF plasma in order to remove PDMS-St wetting of the NW top parts to allow further electrical contacting.

To compare different contacting strategies, we deposited onto the NW top parts (i) Cr/Au/Cr 5/50/20 nm metal layers, (ii) SWCNT film with 40 nm thickness, 80% transparency and 250 Om*cm sheet resistance [Tsap2018], (iii) pristine FPS, and (iv) FPS mixed at 100 : 1 mass proportion with MWCNT with 20 μm and 20 nm average length and width, respectively [Rom2016]. Then the PDMS-St/NW membranes were mechanically peeled from the Si wafer with a razor blade and flipped onto an arbitrary holder. The bottom parts of the NWs were protruding from the membranes, and their surface was not covered with the PDMS-St, therefore after the membrane release the samples were ready for the bottom contact deposition. In order to perform a consistent analysis and facilitate comparison of the contacted membranes, the bottom contact material was chosen the same for all (i-iv) samples. The best candidate is SWCNT contact due to its high elastic properties, conformal coverage of the NWs, high conductivity and transparency for both optical and SEM microscopy. The SWCNTs are also the envisioned contact for the optoelectronic devices, allowing to stack the PDMS-St/NW membranes with each other or different material systems in composite structures due to SWCNT contact transparency. The design of the fabricated samples is shown in Figure 3.

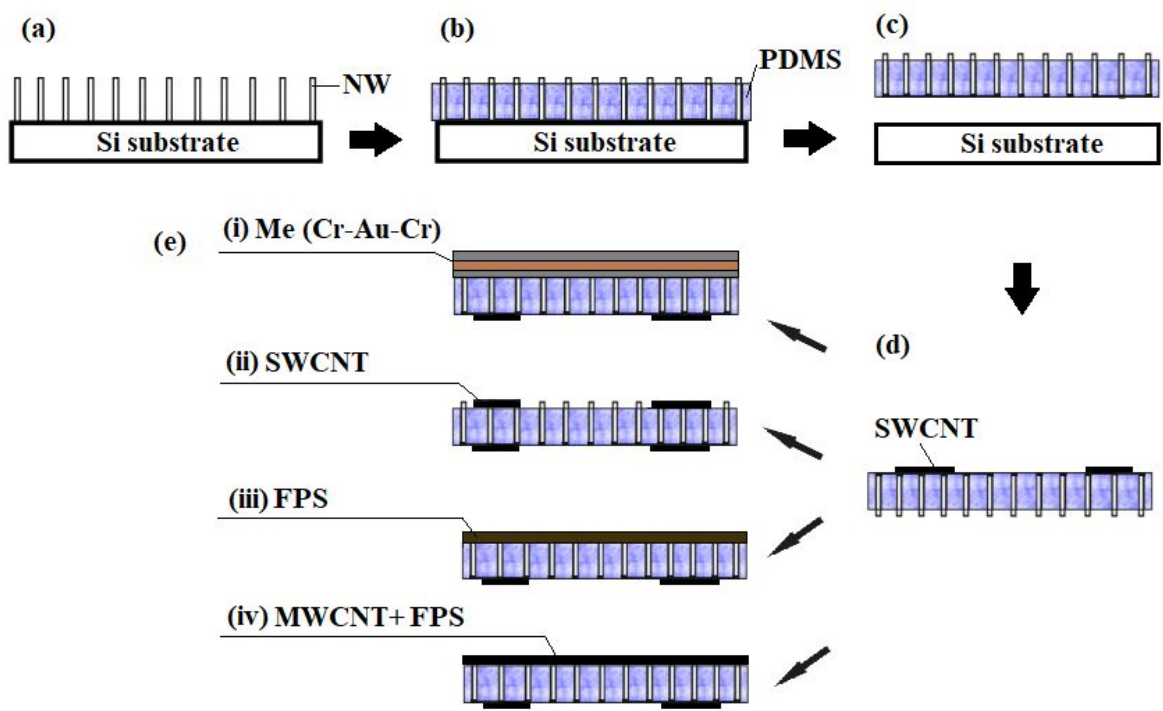

Figure 3. Schematic representation of the fabrication workflow: (a) MBE GaP nanowires on Si substrate; (b) encapsulation of NW array into PDMS-st. membrane; (c) NW/PDMS-st. membrane release; (d) SWCNT bottom contact fabrication to NW bases; (e) top contact fabrication (i)

Cr/Au/Cr metal, (ii) SWCNT pads, (iii) FPS continuous contact, (iv) FPS doped with MWCNT continuous contact.

For the sample (i), the metallic contact Cr/Au/Cr was chosen considering high adhesion of chromium to the NW and PDMS-St material. Au is an excellent material in terms of conductivity and plasticity, but Au adhesion to the PDMS-St is very low. In order to improve adhesion, Cr cladding layer were introduced. The Cr/Au/Cr contact have a desirable Schottky barrier to the n-GaP [Vin1975] material, what allows to distinguish it from shunting and detect proper contacting in electron beam induced current (EBIC) measurements. The metal was deposited with electron-beam physical vapor deposition (for Cr) and thermal evaporation (for Au) using Boc Edwards Auto 500 setup operating at $5 \times 10^{-6}$ mbar.

The detailed procedure for SWCNT synthesis is described elsewhere [mois2006, nas2008, Iak2019]. Briefly, SWCNTs were synthesized by aerosol CVD method in a tubular quartz reactor with floating catalytic bed (T = 880°C; CO as a carbon source, carbon dioxide as a growth promoter, and ferrocene as Fe catalyst precursor). The SWCNTs were collected from the outlet of the reactor on nitrocellulose filter (HAWP, Merck Millipore) for a certain time in order to obtain the desired thickness of thin network. The 40 nm thickness of the SWCNT film was chosen as a material with high conductivity, transparency, and conformal coverage of the NW protruding parts. The SWCNT films on the nitrocellulose filter afterwards can be cut to an appropriate geometry and transferred on the sample without additional manipulations by dry application [Kas2010]. In order to reduce the risk of accidental shunting, the contact pads were fabricated (1 mm² or less in size). The sample (ii) featured similarly fabricated SWCNT on both face and rear surfaces.

The FPS was synthesized by the reaction of catalytic hydrosilylation between PMHS and vinylferrocene according to procedure published in [Der2019] (Figure 4). The molar ratio of the Si–H groups and vinylferrocene was selected so that 50% of the hydride Si–H groups remained unreacted. Vinylferrocene (1.5 g, 7.08 mmol) was added to a benzene solution (10 mL) containing 20 μL of a 0.1 M solution of Karstedt's catalyst in xylene in a tube purged with argon. The mixture was stirred at room temperature for 1 h. Then, a solution of PMHS (849 mg, 14.15 mmol of –OSiHCH$_3$– moieties) in dry benzene (10 mL) was added dropwisely for 1 h. The contents of the tube were sealed and stirred at 40 °C for 24 h. The solvent was removed by rotary evaporation. Yield of FPS: 2.35 g (100%); brown viscous liquid. The obtained FPS was fully characterized with nuclear magnetic resonance (NMR) spectroscopy [Der2019], dissolved in 2 mL dichloromethane (DCM) and applied by drop-casting onto the face surface of the PDMS-St/NW membrane before the release from the Si wafer. The FPS gel requires evaporation of DCM to achieve proper self-cross-linking by the reaction between Si–H groups [Der2019b; Der2019] (Figure 4). After 24 h DCM evaporation inside a ventilated chemical hood, the samples (iii) were put into an oven at 80 °C for 30 min, resulting in a homogeneous cross-linking of the whole volume of the FPS. The fabricated cross-linked FPS contact also serves as a good mechanical support, facilitating the PDMS-St/NW membrane release from the Si wafer.

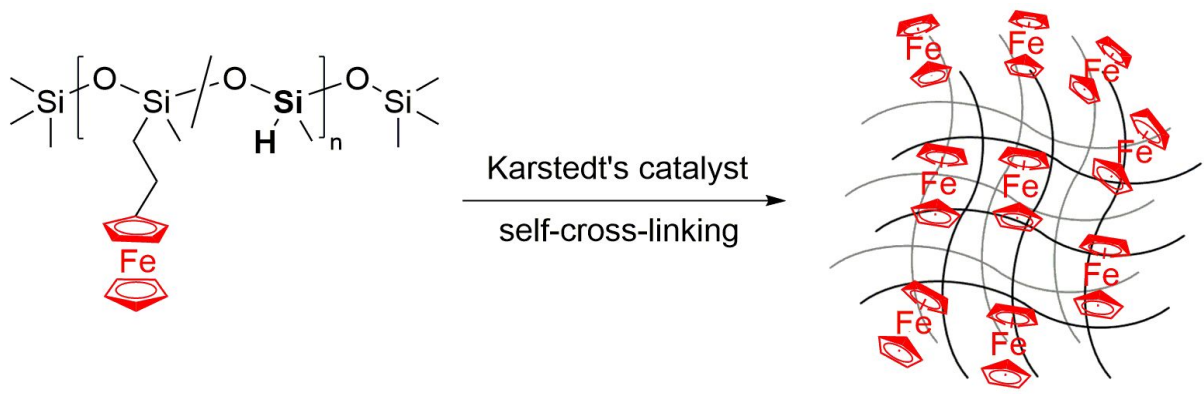

Figure 4. FPS and its self-cross-linking.

To increase the inherent FPS conductivity we prepared samples (iv) similar to (iii), but with the MWCNTs diluted in FPS. While in the air the SWCNTs exhibit higher conductivity and transparency, the PDMS/MWCNT solution demonstrates higher respective properties compared to PDMS/SWCNT [Gul2013] due to less perturbation of the electron transport in solid solution, while SWCNTs partly loses properties of graphene-like 2D material because of charge carrier scattering [Chen2008]. For homogeneous dispersion of MWCNT in FPS the tubes were first dispersed in DCM with 2 mm diameter ultrasound probe at 84 W power for 60 min, and then the MWCNT/DCM solution was added to FPS and ultrasounded at 24 W for 20 min. Then the FPS with MWCNTs was applied to the PDMS-St/NW surface and left in the ventilated hood for 24 hour for DCM evaporation and then baked for 30 min at 80 °C similar to the sample (iii).

After top contact fabrication all (i-iv) samples were processed in a similar way. The samples were peeled with a razor blade, flipped onto an arbitrary holder, i. e. a piece of Si wafer, Al plate or glass, and the SWCNT contact pads of average size approx. 1 mm² were applied. Due to mechanical instability and advanced chemistry the samples were controlled at each step by optical and electron microscopy.

To allow stable I-V measurements the samples were put with the NW top parts onto a Si wafer with a $Au_{0.85}Ge_{0.15}$ 200 nm conductive layer serving as the bottom electrode. Then small droplets of silver lacquer CDS Electronique L-200 was put on the SWCNT contact pads to allow probe tip connection. The high quality silver lacquer was chosen instead of standard paste or paint because of fast drying, i. e. 10 min., low viscosity facilitating small droplet application, low charging and outgassing inside SEM setup, high adhesion to SWCNTs, and traceless removal in acetone.

In order to prove electrical connection between the NW array embedded into PDMS-St membrane and the different contacting layers, EBIC measurements were performed. It was expected that all contacts to (i-iv) samples should demonstrate electric barrier to the n-GaP material, therefore EBIC mapping allows visualisation of the intrinsic electric fields, provided that the NWs are electrically connected. EBIC experiment was performed in Supra 40VP Zeiss SEM setup equipped with Gatan EBIC hard and software coupled with Stanford SR570 sourcemeter, and Kleindiek NC30 micromanipulators.

Finally, we fabricated p-GaP : Be PDMS-St/NW membrane sample similar to n-GaP sample (ii), i. e. with SWCNT contact pads on both sides. The I-V characteristics for the p-GaP membrane were measured in order to define the ohmicity of the contact of SWCNT to p-GaP, which was expected to have low barrier or even to be ohmic due to the hole conductivity of the SWCNT [Sun2011].

4. Results and discussion

SEM microscopy of n-GaP : Si NWs demonstrated 10 µm and 130 nm average height and width, respectively (Figure 5).

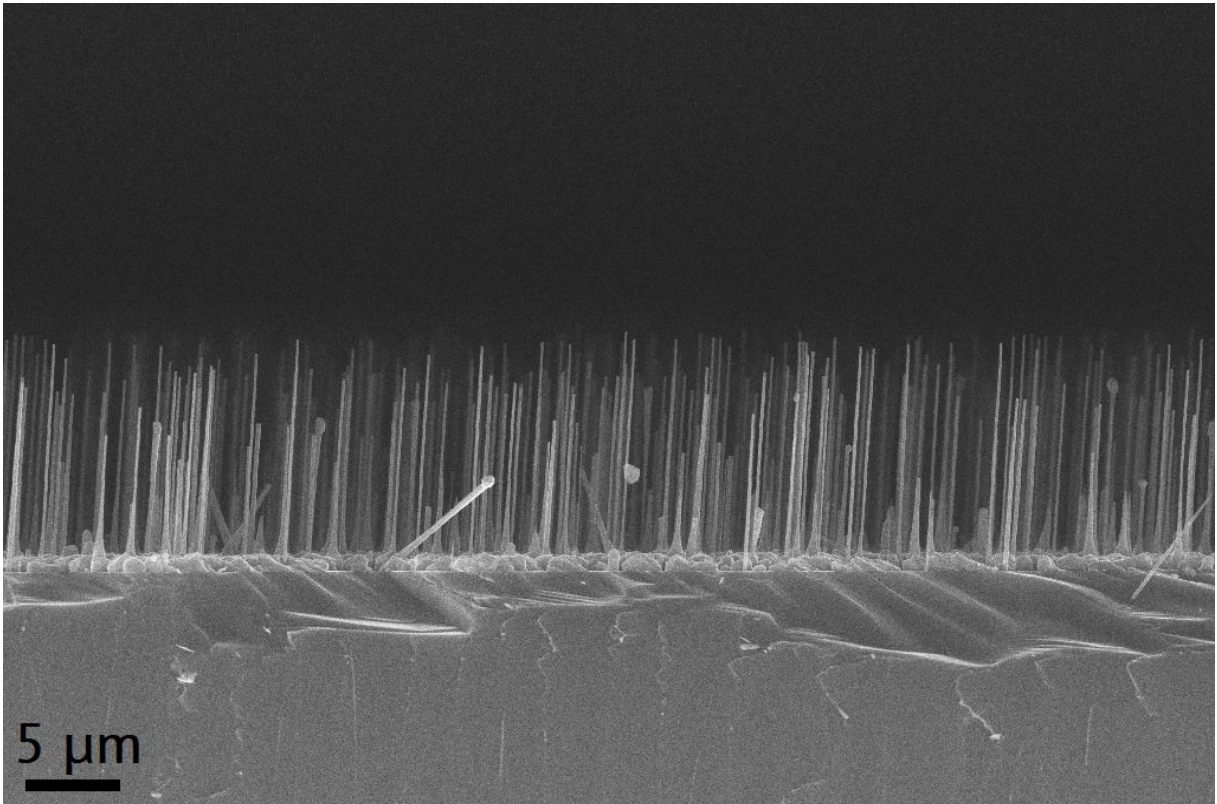

Figure 5. SEM image of the cut edge of the as-grown n-GaP : Si NWs on Si wafer.

The NW array demonstrated significantly inhomogeneous NW height distribution typical for self- induced Ga-catalyzed growth. Therefore, after encapsulation in PDMS-St the lower NWs were buried and only high enough NWs were revealed after plasma etching (Figures 6, 7 and 8). The average density of the contacted NW array is estimated to be 0.005 NW/µm².

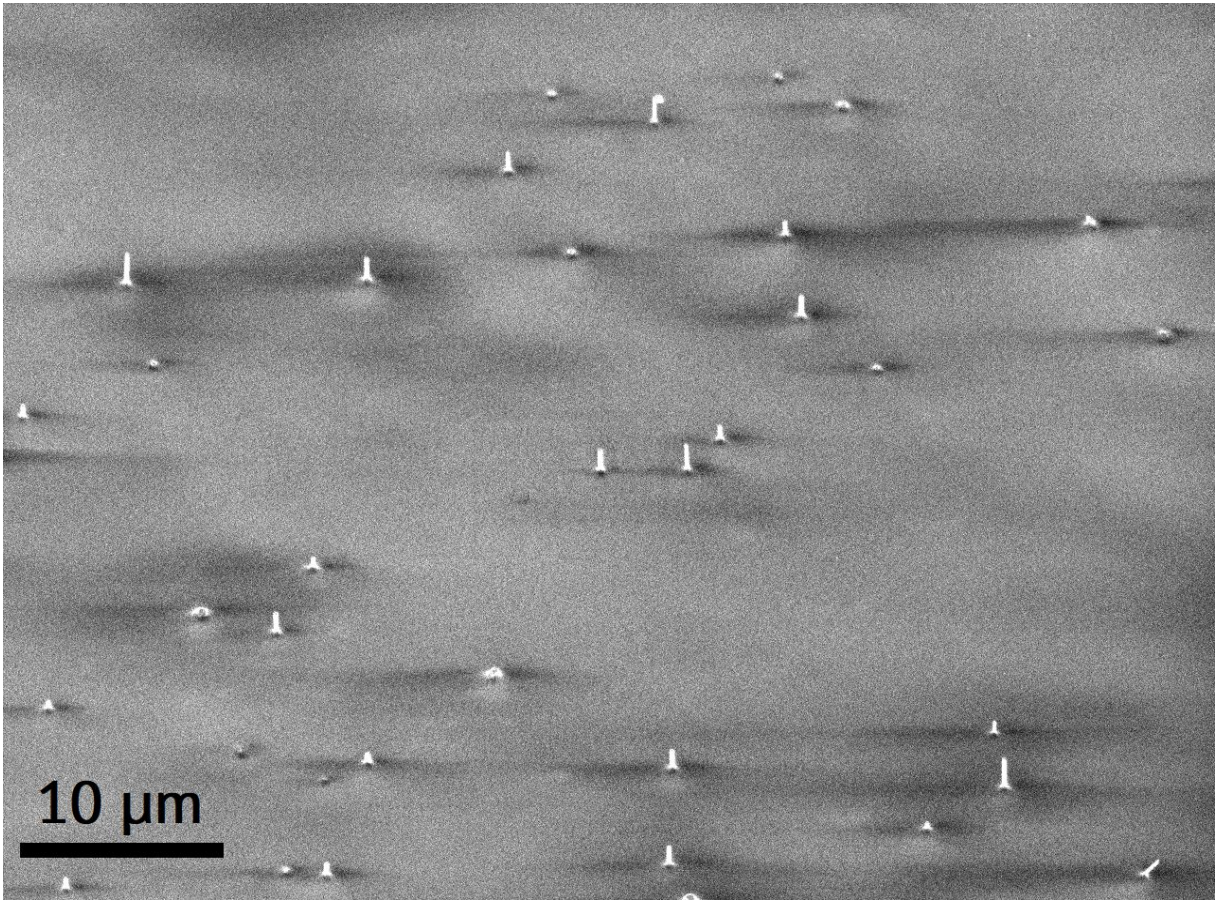

Figure 6. 45° angle SEM image of n-GaP : Si NW array encapsulated into PDMS-St membrane.

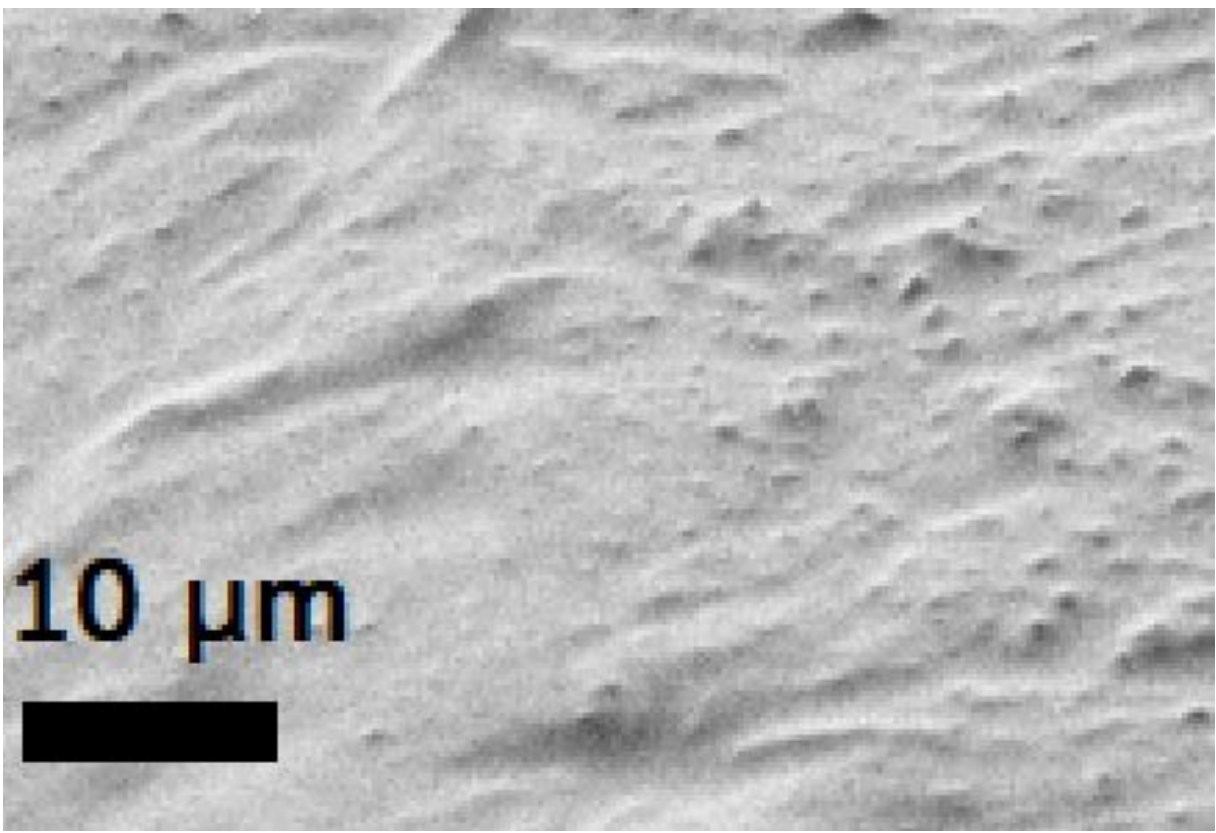

Figure 7. SEM image of Cr/Au/Cr contact on the top side of the n-GaP PDMS-St/NW membrane (sample (i)).

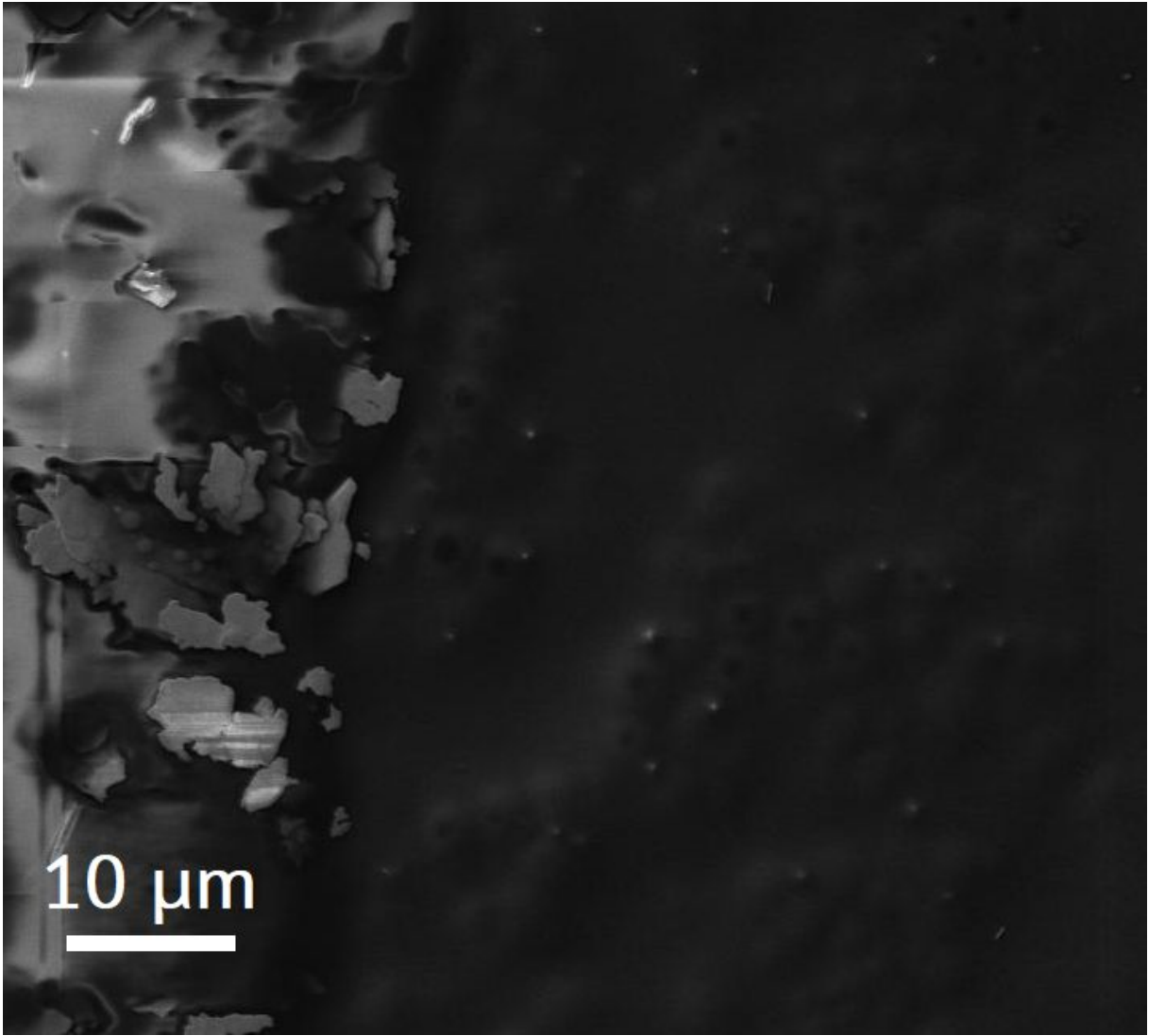

Figure 8. SEM image of SWCNT contact pad with silver lacquer droplet (left) on the top side of n-GaP : Si NW/PDMS-St membrane (sample (ii)), bright contrast dots in the right area correspond to contacted with SWCNTs NWs.

After the membrane release, the Si substrates were studied with SEM to control NWs transfer to the PDMS-St. The SEM images showed no vertical NWs (Figure 9), the chaotically laying NWs can be attributed to the parasitic non-vertical NW growth.

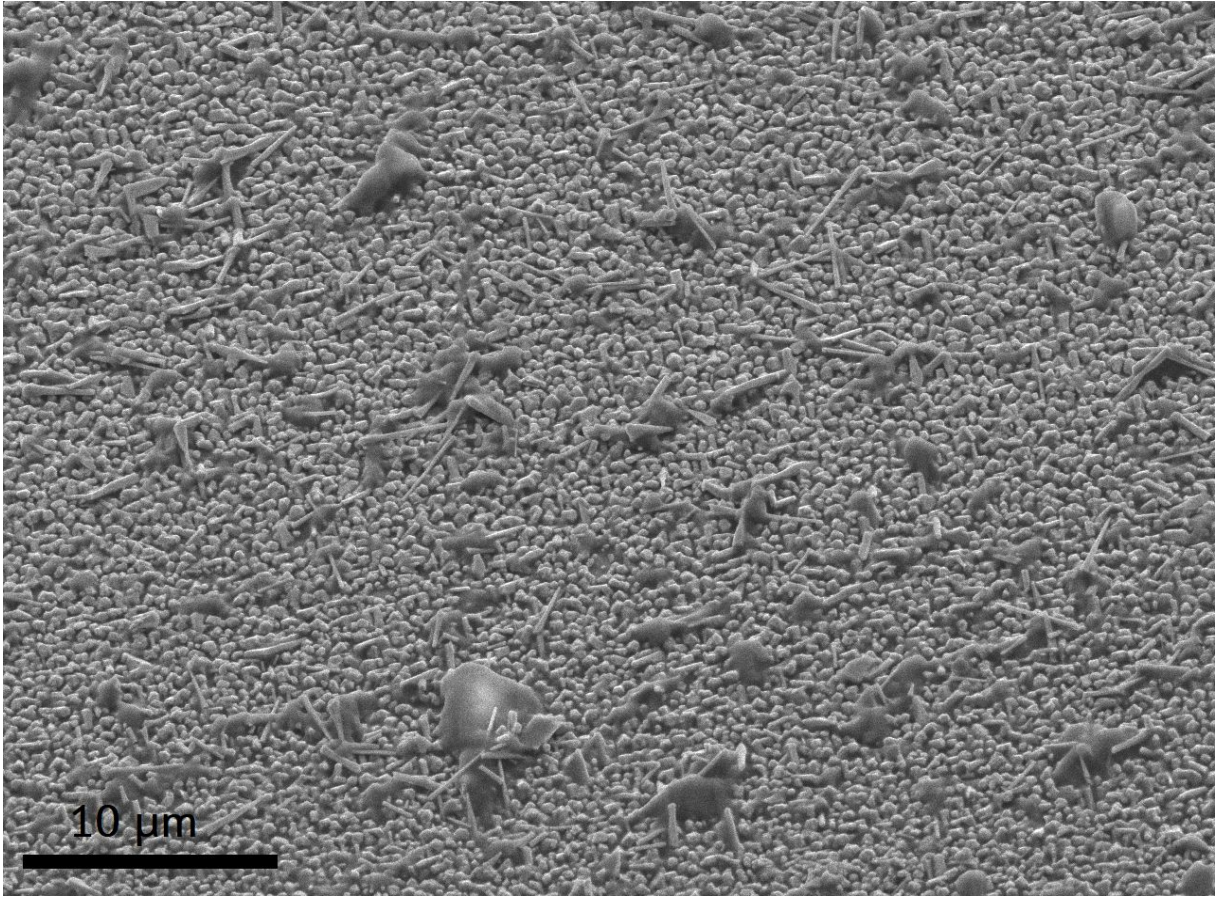

Figure 9. 45° tilted SEM image of the Si substrate of n-GaP : Si NW sample after the membrane release.

Further in depth n- and p-GaP NWs structural characterization will be published elsewhere. Current-voltage characteristics of fabricated samples (i-iv) were measured (Figure 10). Because all the contacts are expected to have electric barrier to the n-GaP material, the strongly nonlinear behaviour of the I-V curves may be associated with the reversed Schottky barrier characteristic [Cow1966]. The sample (i) demonstrated lower knee bias in comparison to other samples, which we associate with high surface state density at the interface of Cr/n-GaP NW. The positive voltage branch of the sample (i) and symmetric curve of the sample (ii) demonstrated knee value at 5 V, which we attribute to reversed current in SWCNT/n-GaP Schottky barrier. The sample (iii) demonstrated instabilities of the current at positive applied bias, which may originate from the mechanical instability or the piezo or thermal striction of the PDMS-St membrane material, similar instabilities at I-V curve, accompanied by EL blinking, were also observed for LED membrane devices [Nep2016]. The sample (iii) demonstrated a high knee voltage of 7-8 V and relatively low current. Sample (iv) demonstrated an I-V curve shape similar to sample (ii), but less current and higher knee voltage. The FPS and FPS/MWCNT contacts demonstrated a high mechanical stability. The calculated current through individual wires and the derived current density of the measured samples are presented in Table 1.

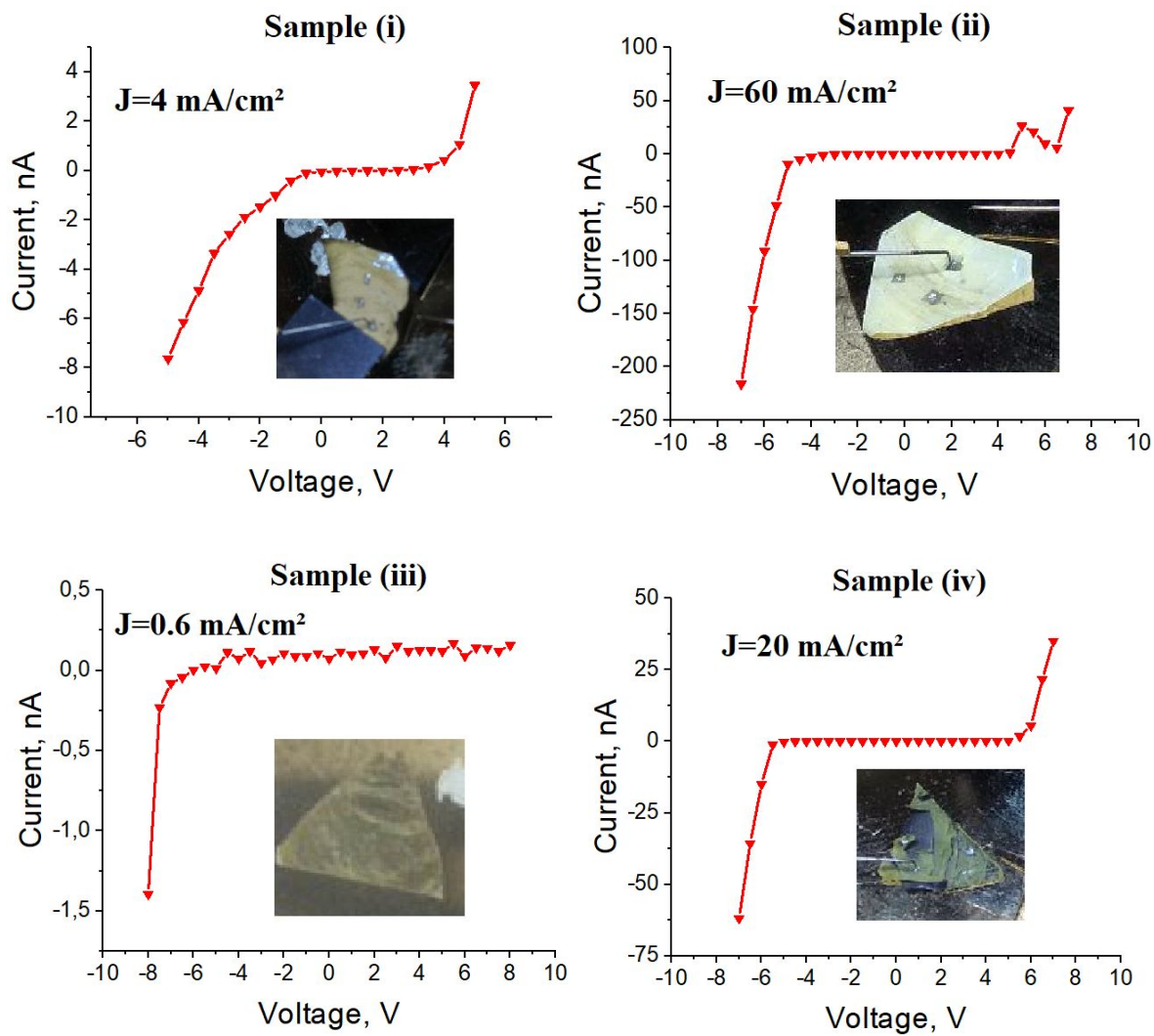

Figure 10. Current-voltage curves of n-GaP : Si samples (i-iv). The photographs in the insets demonstrates the corresponding PDMS-St/NW membrane (grayish yellow), and three SWCNT contact pads (gray squares); the size of contact pads is about 0.25 mm². Samples have similar SWCNT contact pads on the NW bases, and different top contacts: (i) Cr/Au/Cr metallic contact, (ii) SWCNT contact pads, (iii) continuous FPS contact, (iv) continuous FPS/MWCNT contact.

In order to prove that the measured current originates from NW electrical connection and not shunting through possible cracks in the PDMS-St membrane, a control non-doped GaP NW sample with both side SWCNT contact was processed. The morphology of the control sample is very similar to n-GaP NWs, Figures 5 and 11, respectively, so an accurate comparison is possible. The sample was processed similar to the best performance sample (ii), the measured I-V curve is presented in Figure 12. The signal is below the noise level, the -150 pA current at the turning on is a typical Keithley 2400 error due to the electrostatic discharge. We conclude that the control NWs are not conductive, and the n-GaP NW/PDMS-St membranes analyzed above demonstrated current through the NWs and not through possible shunting.

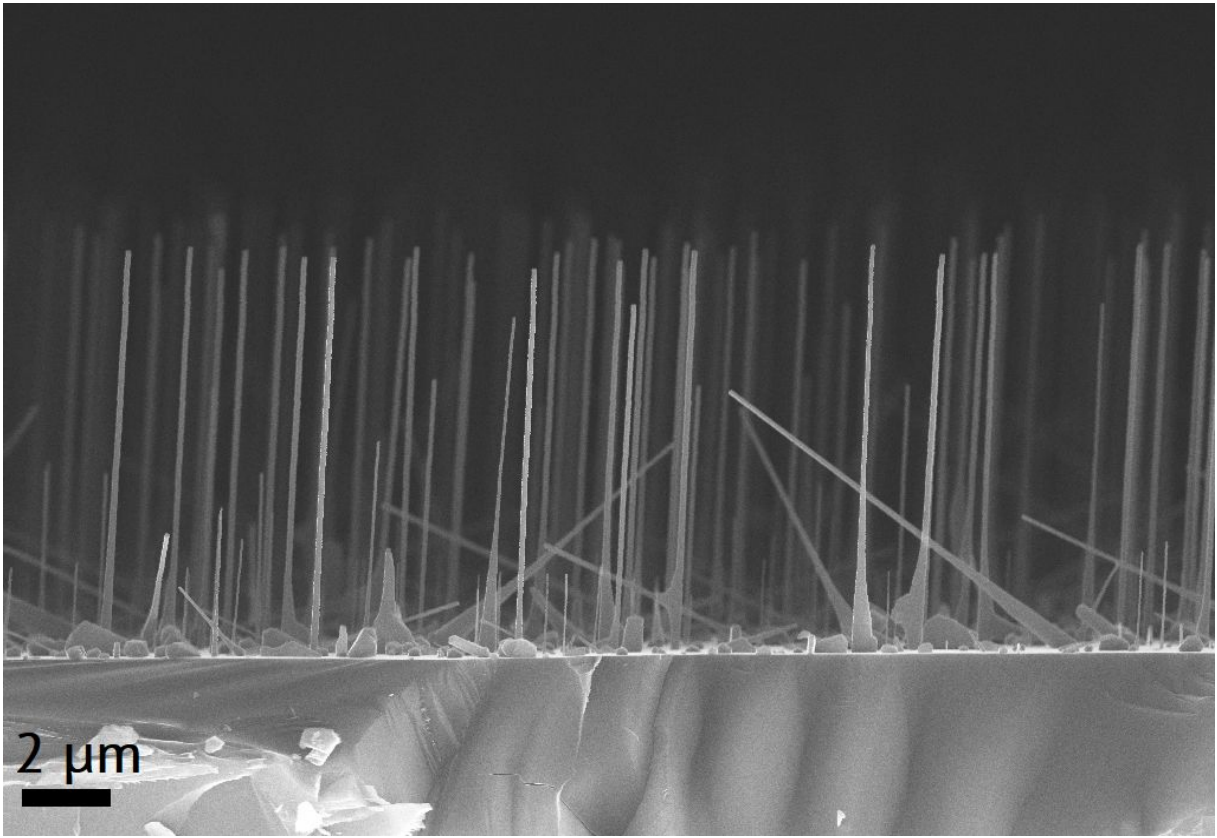

Figure 11. SEM image of the control non-doped GaP NW sample with similar to n-GaP NW sample morphology.

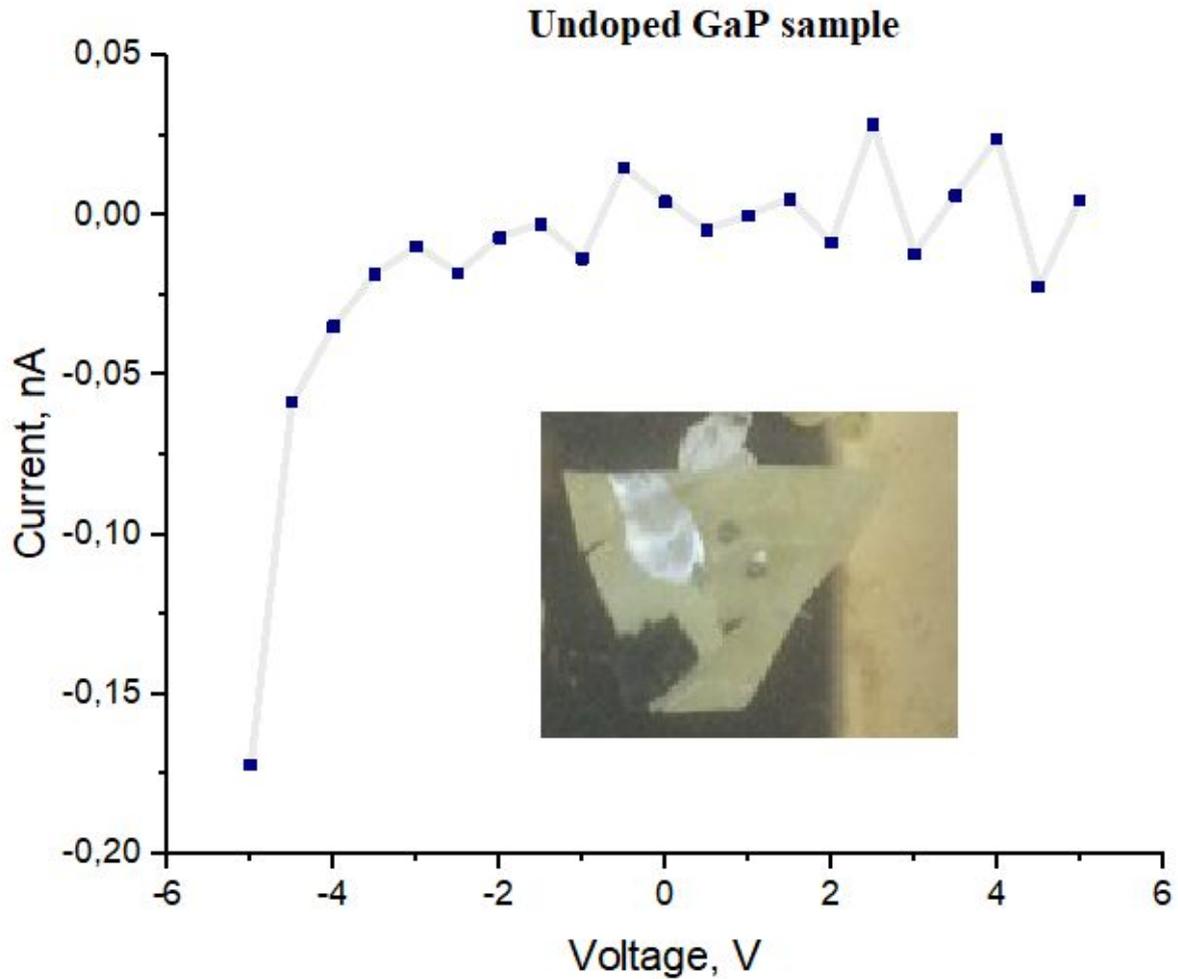

Figure 12. The measured I-V curve of the control non-doped GaP NW sample. The photograph in the inset demonstrates the PDMS-St/NW membrane (grayish yellow), and three SWCNT contact pads (gray squares); the size of contact pads is about 0.25 mm².

To supply the main n-GaP : Si NW sample series of the presented work with additional comparison, we processed p-GaP : Be NW similar to samples (i-ii). The p-GaP NW sample has different morphology (Figure 13), namely 3 µm and 250 nm average NW height and width, respectively. The average height and width are 3 µm and 250 nm, respectively. The NWs feature a Ga metallic droplets on top of them, which remain after the epitaxial growth interruption.

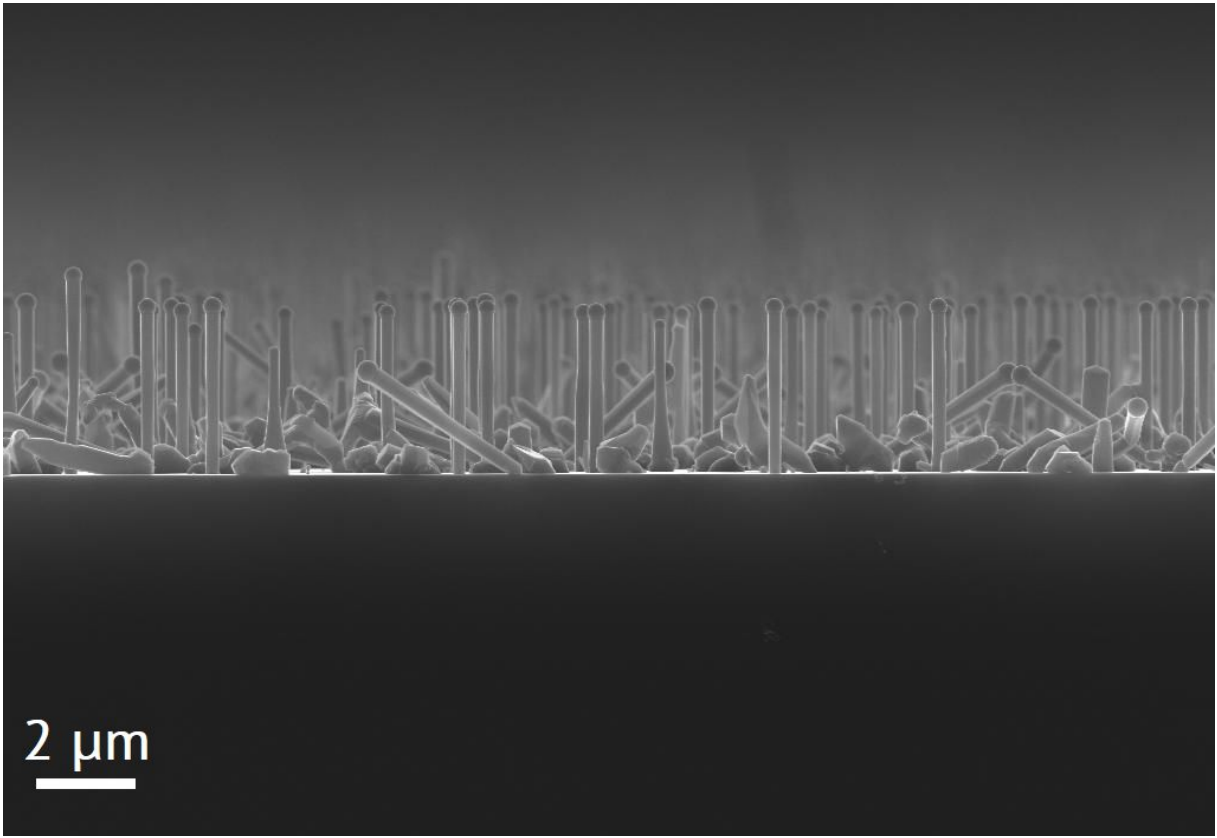

Figure 13. SEM image of the cut edge of the as-grown p-GaP : Be NWs on Si wafer.

The NW array has higher homogeneity of NW height distribution because the Ga droplets were not completely spent. After PDMS-St encapsulation the revealed NW array is presented in Figure 14. The estimated NW density is 0.0125 NW/µm².

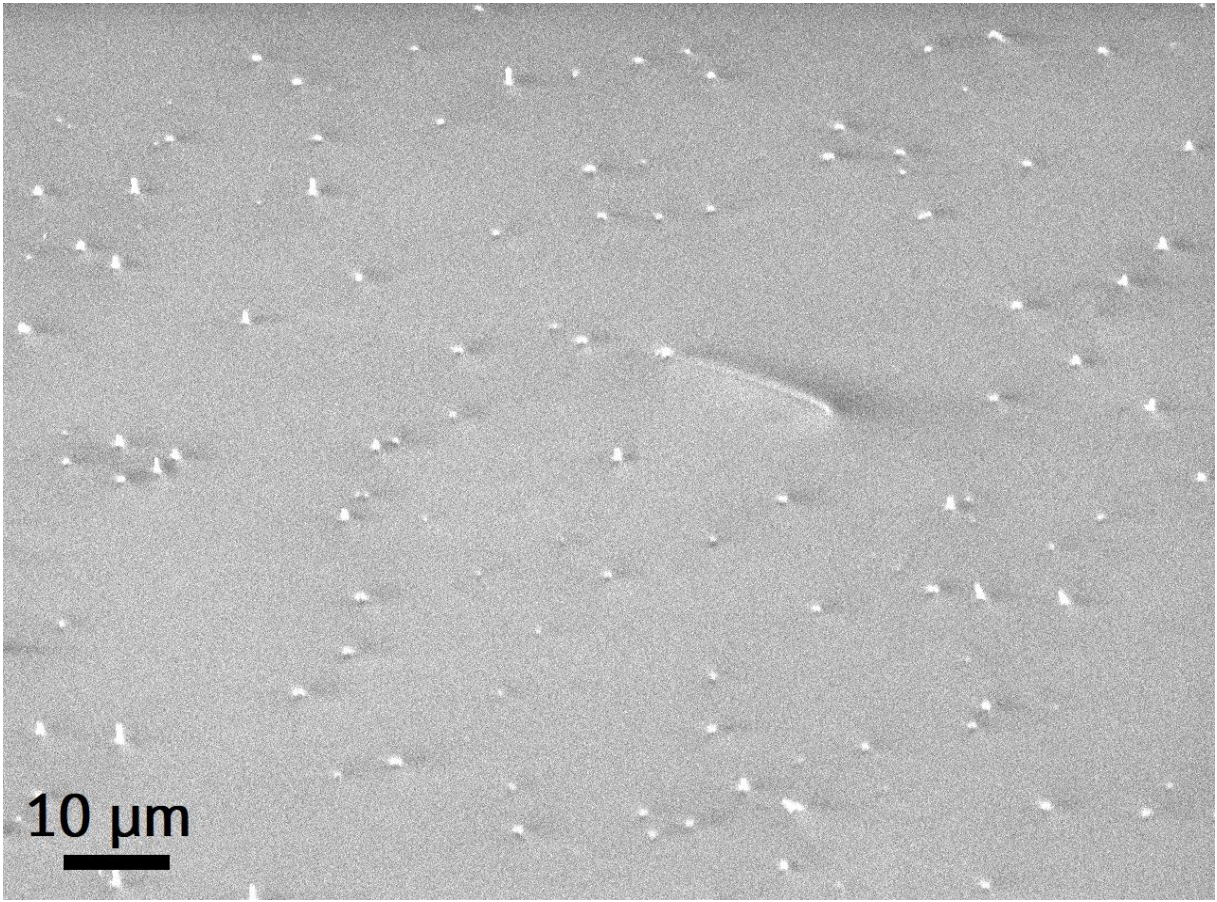

Figure 14. 45° angle SEM image of p-GaP : Be NW array encapsulated into PDMS-St membrane.

The measured I-V curves are presented in Figure 15. The sample with metallic contact demonstrated higher resistivity in comparison to the SWCNT sample, the knee voltage is similar for both samples and is estimated to be in the range of 0.5-1 V. We attribute the non-linearity of the I-V curve to the Ga droplets, having small Schottky barriers with the p-GaP material and/or the SWCNT contact. The derived current values are presented in Table 1.

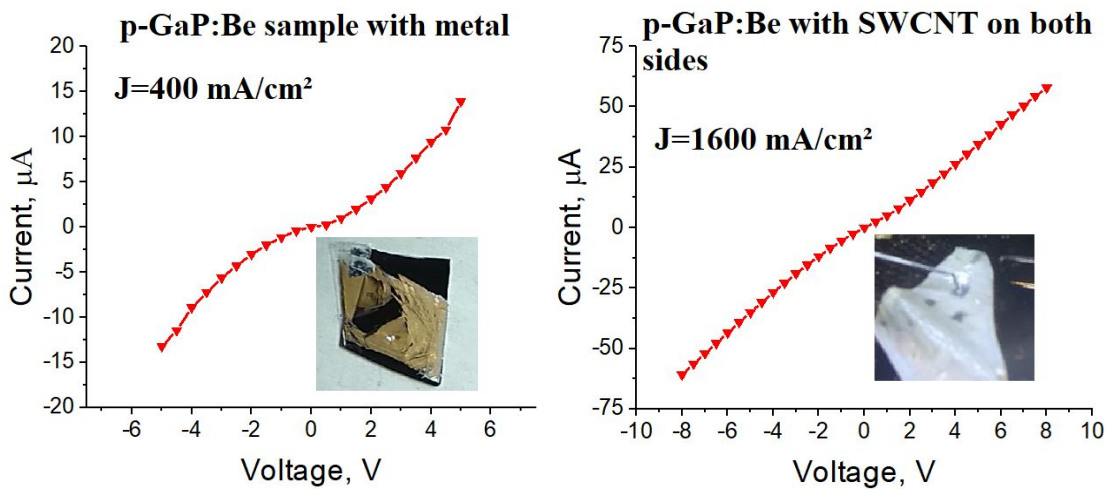

Figure 15. The measured I-V curves of p-GaP : Be NWs in PDMS-St/NW membrane with Cr/Au/Cr (left) and SWCNT (right) top contact. The bottom contact of both samples is SWCNT contact pads. The inset photos show the measured samples.

Table 1. Calculated current through individual NWs and corresponding current densities.

| Sample | Current at 1 V above opening, pA/NW | Current density, mA/cm² |
|---|---|---|
| (i) n-GaP Cr/Au/Cr SWCNT | 2 | 4 |
| (ii) n-GaP SWCNT both sides | 30 | 60 |
| (iii) n-GaP FPS SWCNT | 0.3 | 0.6 |
| (iv) n-GaP FPS/MWCNT SWCNT | 10 | 20 |
| p-GaP Cr/Au/Cr SWCNT | 800 | 400 |
| p-GaP SWCNT both sides | 3200 | 1600 |

The Schottky barrier build-in field was tested with the EBIC microscopy [Lym2012]. The EBIC measurements for PDMS-St/NW membranes is very challenging mainly due to mechanical instability, charging and low magnitude of the signal. However, for the reported work low resolution EBIC is sufficient to prove the electrical connection of the NW array and the existence of contact barriers, provided the noise level allows to distinguish EBIC signal and attribute it to NW location sites. The Figures 16 and 17 demonstrates SEM images and EBIC maps of the samples (i-ii). The signal resolution allowed to attribute the EBIC signal contrast to the NW cites in SEM image. The signal amplitude is low due to the operation in the reversed diode characteristic curve, however, it allows to detect the presence of a Schottky barrier, i. e. the electrical connection to the NWs.

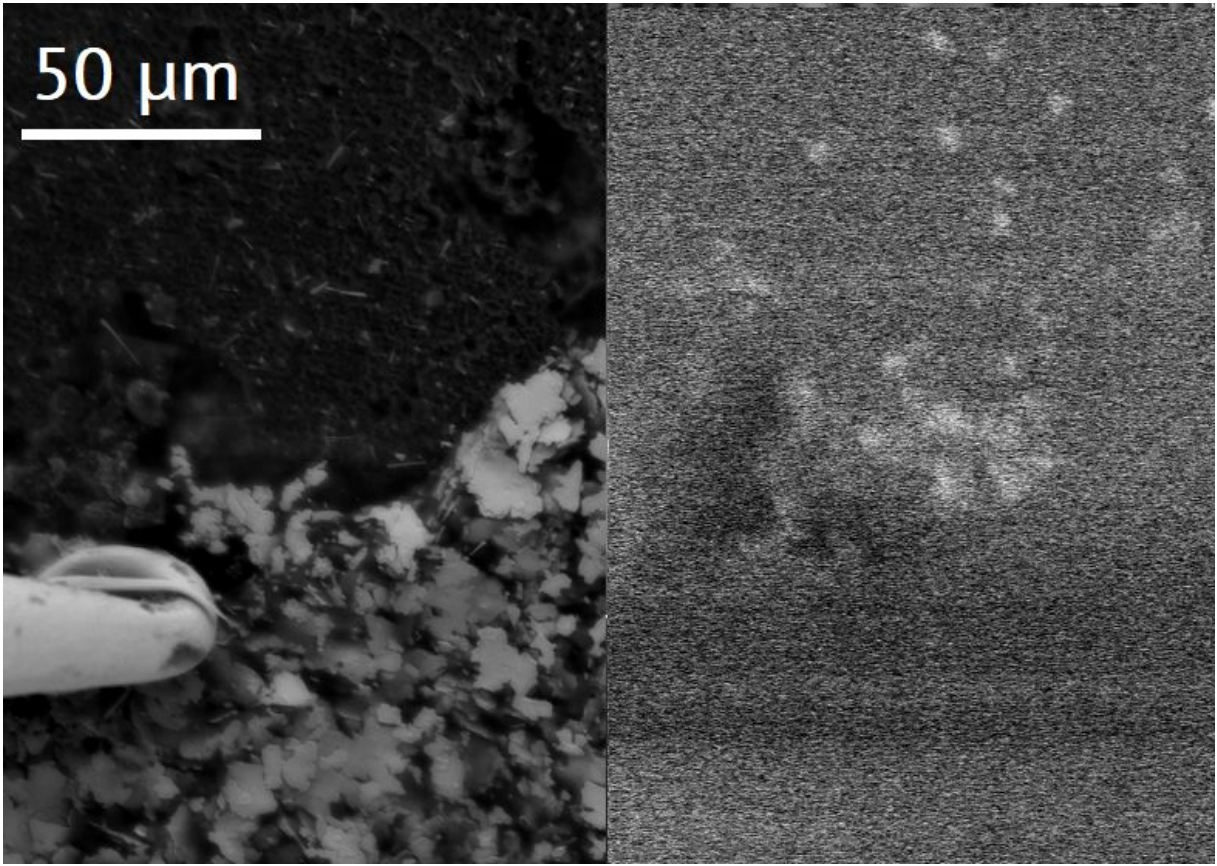

Figure 16. SEM image (left) and EBIC map (right) of the sample (i) with metallic contact, bright contrast of EBIC map is attributed to electrically connected wires. 10 kV e-beam acceleration voltage.

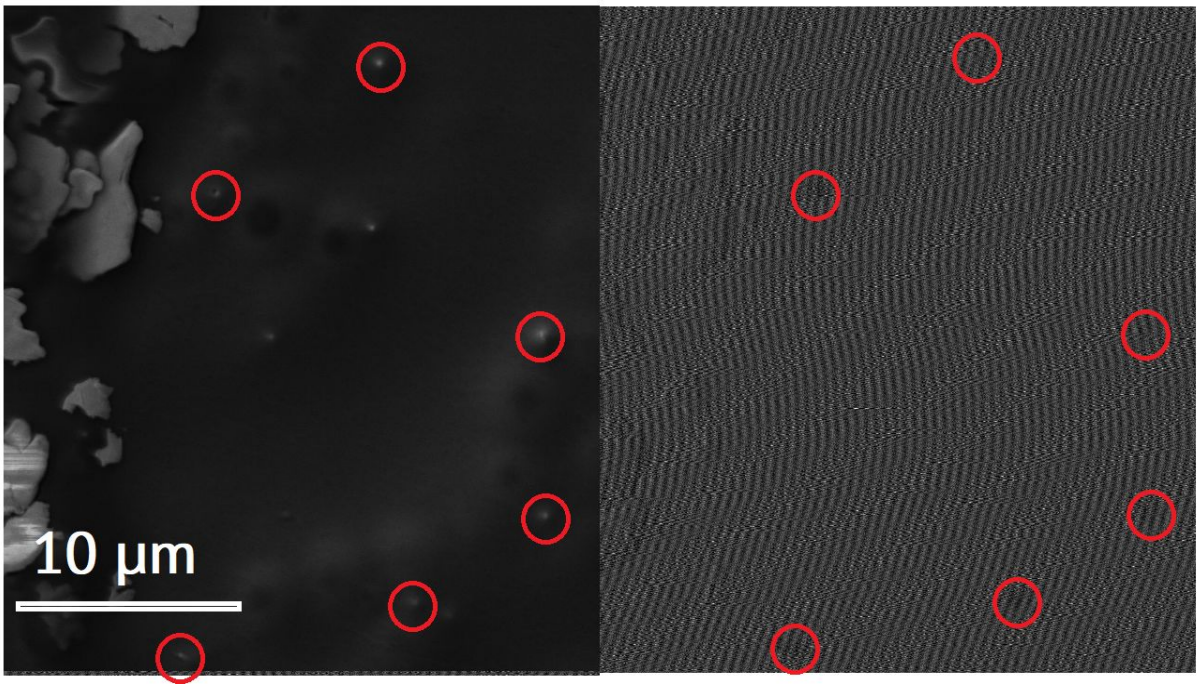

Figure 17. SEM image (left) and EBIC map (right) of the sample (ii) with both side SWCNT contact, dark contrast of EBIC map in red circles is attributed to electrically connected wires. 5 kV e-beam acceleration voltage.

5. Conclusions

The developed PDMS-St allowed to encapsulate and release cm² size NW membranes. The developed G-coating method was used to produce 10 and 3 μm thick suspended NW/PDMS-St membranes. The fabricated n- and p-GaP NW/PDMS-St membranes were processed with Schottky and near-ohmic contacts, respectively. The developed FPS successfully served as flexible semitransparent contact, MWCNT doping efficiently increased the FPS contact performance. The SWCNT network shown its quality as a transparent electrode with high stability and nearly ohmic contact to p-GaP : Be. The electrical connection of the NW array in the PDMS-St membrane was proven by control samples processing and EBIC microscopy.

The presented work proposes advanced chemistry, membrane, and contact fabrication techniques, which can be adapted for fabrication of NW optoelectronic devices.

Acknowledgements

The Russian Science Foundation is greatly acknowledged for financial support for NWs growth (V. F. thanks Agreement No. 18-72-00231) and for CNTs synthesis (Agreement No. 17-19-01787). Synthesis of FPS and its nanocomposites creating were funded by the Russian Foundation for Basic Research (project No. 18-33-20062_mol_a_ved for RMI). The authors from Skolkovo Institute of Science and Technology acknowledge MIT Skoltech Next Generation Project. M. T. acknowledges the financial support from EU ERC project "NanoHarvest" (grant No. 639052). The authors also acknowledge the Interdisciplinary resource center for nanotechnology of Research park of Saint Petersburg state university.